\documentclass[aps,prx,twocolumn,superscriptaddress,floatfix]{revtex4-2}

\usepackage[T1]{fontenc}
\usepackage{amsmath,amssymb,bm}
\usepackage{graphicx}
\usepackage{xcolor}
\usepackage{hyperref}

\newcommand{\avg}[1]{\left\langle #1\right\rangle}

\begin{document}

\title{Programmable dipolar interaction geometry selects stripe-family order in a molecular lattice quantum simulator}

\author{Chao Zhang}
\email{chaozhang@ahnu.edu.cn}
\affiliation{Department of Physics, Anhui Normal University, Wuhu, Anhui 241000, China}

\begin{abstract}
Microwave-dressed polar molecules offer a route to lattice quantum simulators in which the angular form of long-range dipolar interactions, not only their overall strength, can be engineered.
We study this setting in a minimal hard-core Bose lattice model on a square optical lattice, with particles interacting through a sign-changing non-axisymmetric dipolar tail \(\mathcal V(\mathbf r)\propto (x^2-y^2)/(x^2+y^2)^{5/2}\) that is repulsive along one lattice axis and attractive along the other.
Using worm-algorithm path-integral quantum Monte Carlo simulations, supported by a hard-core spin mapping and a Gutzwiller soft-mode diagnostic, we find two regimes controlled by \(t/V\): at larger \(t/V\) the system remains superfluid but develops a pronounced directional stiffness anisotropy, while at smaller \(t/V\) it forms a stripe solid selected in the \((q,0)\) axial family, corresponding to real-space stripes parallel to \(y\).
The leading ordering wave vector remains in this axial family but reorganizes with filling, showing that the robust ordered object is a family of stripe states rather than one fixed commensurate Bragg peak.
Near the closure of the stripe lobe, averaged observables can mimic a narrow supersolid signal; measurement-resolved stripe structure-factor histograms instead reveal first-order switching between superfluid and stripe-solid sectors.
NaCs lattice estimates place the relevant \(V/t\) window within reach of modest effective dressed dipole moments, linking the predicted stripe-family order and its experimental diagnostics to accessible molecular quantum-simulation scales.
\end{abstract}

\maketitle

\section{Introduction}
\label{sec:intro}

Ultracold polar molecules are emerging as quantum simulators in which long-lived internal states and strong dipolar interactions can be combined with optical lattices or tweezer arrays to realize programmable many-body Hamiltonians~\cite{MicheliNatPhys2006,GorshkovPRL2008}.
Recent progress has sharpened this opportunity: shielding and dressing protocols now control collisional loss, elastic scattering, Bose condensation, and strongly correlated molecular phases in continuum gases~\cite{KarmanPRL2018,AndereggScience2021,SchindewolfNature2022,LinPRX2023,BigagliNatPhys2023,BigagliNature2024,KarmanPRXQ2025,WeiZhangPRXQ2025,JinPRL2025,ZhangNature2026}.
A natural next step is to place such engineered molecular interactions in an optical lattice, where their competition with tunneling and filling realizes an extended Hubbard problem with a programmable interaction profile.

Optical lattices provide controlled realizations of Hubbard-type models in which geometry, dimensionality, tunneling, and interactions can be tuned with high precision~\cite{BlochRMP2008}.
With onsite interactions alone, the equilibrium Bose--Hubbard model is governed by the familiar competition between superfluid and Mott insulator phase.
Longer-ranged interactions enrich this problem by allowing density-wave order, supersolidity, and ordering wave vectors set by nonlocal couplings~\cite{BaranovPhysRep2008,LahayeRPP2009,GoralPRL2002,CapogrossoPRL2010,PolletPRL2010,TrefzgerJPB2011,ZhangNJP2015TiltedDipolar}.
Dipolar atoms and molecules provide a well-established route to such extended Hubbard physics: their long-ranged and anisotropic interactions make the density-ordering channel sensitive to the full interaction profile~\cite{GoralPRL2002,MenottiPRL2007,BuchlerPRL2007,CapogrossoPRL2010,DanshitaPRL2009,BandyopadhyayPRA2019,MicheliPRA2007,ZhangPRA2018SiteDilutedPolarMolecules}.
Much of the lattice dipolar literature starts from interactions that are isotropic in the lattice plane or from conventional dipoles polarized perpendicular to, or tilted with respect to, the plane~\cite{CapogrossoPRL2010,TrefzgerJPB2011,BandyopadhyayPRA2019,AleksandrovaPRA2024,ZhangPRA2021TiltedDipolar,MaciaPRA2014,ZhangPRA2022Tilted3D}.
Those settings already support checkerboard solids, supersolids, and stripe solids, with disorder providing another route to glassy or fragmented density-ordering behavior~\cite{ZhangPRA2018DisorderedDipolar}.
Here we instead focus on a different interaction geometry that is natural from the perspective of programmable molecular dressing: the leading in-plane harmonic is explicitly \(d_{x^2-y^2}\)-like and changes sign between the two lattice axes, so axial and diagonal density modulations are distinguished already at the two-body level~\cite{KarmanPRXQ2025,DengNatCommun2025,ZhangNature2026}.

Recent experimental and theoretical progress in continuum molecular gases has moved beyond simple fixed-dipole interactions by using microwave and electric fields to suppress loss, tune scattering, and engineer dressed long-range potentials, making the complementary lattice question timely.
Microwave and electric-field shielding protocols have progressed from theoretical proposals to experimental control of collisional loss and elastic interactions~\cite{GorshkovPRL2008,KarmanPRL2018,MatsudaScience2020,AndereggScience2021,SchindewolfNature2022,LinPRX2023,MukherjeeHutsonPRR2024,DuttaPRR2025}.
In particular, recent NaCs experiments have established a dipolar ground-state gas, a collisionally stable bosonic molecular gas, Bose--Einstein condensation, dressed-state spectroscopy with magic trapping, few-body loss diagnostics, and self-bound strongly dipolar molecular droplets~\cite{StevensonPRL2023,BigagliNatPhys2023,BigagliNature2024,ZhangPRL2024Magic,StevensonPRL2024,ZhangNature2026}.
Field-linked resonances and microwave shielding show that molecular scattering channels and short-range losses can be controlled, while polarization-structured microwave dressing provides a route to modifying the angular form of the effective long-range interaction~\cite{ChenNature2023,ChenNature2024Tetramer,KarmanPRXQ2025,DengNatCommun2025}.
Motivated by this direction, we use the \(d_{x^2-y^2}\)-like interaction below as a minimal hard-core lattice benchmark for engineered non-axisymmetric molecular interactions; a protocol-specific Wannier projection is discussed separately in Sec.~\ref{sec:implementation}.

These molecular advances connect to a broader continuum effort in which anisotropy, roton softening, shielding cores, and self-organization generate supersolidity, vortex textures, droplets, self-bound gases, and other directional patterns in dipolar gases~\cite{LahayeRPP2009,MaciaPRL2012,MaciaPRA2014,LuPRL2015,BombinPRL2017,NorciaNature2021,BlandPRL2022,KlausNatPhys2022,RecatiNatRevPhys2023,CasottiNature2024,SchmidtPRR2022,LangenPRL2025,CiardiPRL2025,WeiZhangPRXQ2025}.
A complementary lattice direction is to place engineered molecular interactions in an optical lattice: the lattice depth sets the hopping amplitude \(t\), dressing fields set the interaction scale and, in suitable schemes, its angular profile, while the total molecule number and trap determine the range of local fillings.
Related lattice studies with high-finesse cavities show that additional nonlocal optical modes can further reshape dipolar-boson ordering and drive thermally assisted crystallization~\cite{HebibZhangPRA2023CavityDipolar,HebibZhangPRB2024Thermocrystallization}.
Here we ask how a comparable sign-changing angular interaction is reorganized by a square lattice, where discrete reciprocal-lattice channels compete and the resulting density order can be diagnosed through momentum-resolved density correlations.

This raises a concrete lattice many-body question: when the interaction itself carries a built-in \(d\)-wave-like sign structure, which density-ordering channel is selected, how does that tendency compete with superfluid coherence, and how should the ordered regime be characterized when the preferred wave vector changes with filling?
The key issue is not simply whether stripes appear, but whether the interaction selects an entire directional family of ordering vectors rather than a single fixed commensurate crystal.

In this work, we address these questions in a representative hard-core lattice setting.
We study a square-lattice hard-core Bose--Hubbard model with a long-range sign-changing interaction \(\mathcal V(\mathbf r)\propto (x^2-y^2)/(x^2+y^2)^{5/2}\), the natural lattice projection of an in-plane \(d_{x^2-y^2}\)-type \(1/r^3\) interaction.
The hard-core constraint represents the large-onsite-repulsion limit of a deep molecular lattice, while the long-range tail retains the nontrivial angular structure of the engineered molecular interaction.
This limiting case keeps the lattice tunability and angular dipolar physics relevant to microwave-dressed molecules, while isolating the many-body consequences of the non-axisymmetric tail.
It therefore provides a clean benchmark for geometry-selected ordering and for resolving apparent coexistence signals near the stripe boundary.

Using path-integral quantum Monte Carlo simulations with worm algorithm~\cite{ProkofevPLA1998,ProkofevJETP1998}, we show that the non-axisymmetric interaction produces two closely related many-body effects.
At larger \(t/V\), where hopping is relatively more important, it deforms the superfluid into a directionally anisotropic state with \(\rho_{s,x}\neq \rho_{s,y}\).
At smaller \(t/V\), where the interaction is relatively stronger, it stabilizes an extended stripe-ordered regime selected in the \((q,0)\) channel, i.e., density wave vectors with finite \(q_x\) and \(q_y=0\), corresponding to real-space stripes running parallel to \(y\).
Moreover, the ordered phase is not best described as a single commensurate crystal with one immutable ordering vector.
Instead, the dominant ordering vector drifts with filling while remaining inside the same stripe family, so the natural diagnostic is a family-resolved Fourier analysis rather than the identification of one isolated Bragg peak.
Near the stripe-lobe closure, this distinction becomes experimentally important: averaged observables can show apparent coexistence, while measurement-resolved stripe structure-factor histograms reveal switching rather than a stable hard-core supersolid.
Before turning to the numerical results, we also develop a short physical picture in terms of a hard-core spin mapping and the momentum dependence of the interaction kernel.
This provides a simple organizing principle for axial family selection.

The paper is organized as follows.
Section~\ref{sec:model} introduces the lattice Hamiltonian, the non-axisymmetric interaction, and the numerical diagnostics.
Section~\ref{sec:analytic} gives a compact physical picture for stripe-family selection and particle-hole symmetry.
Section~\ref{sec:results} presents the ground-state phase diagram, the first-order lobe closure, anisotropic superfluid response, and stripe-family order.
Section~\ref{sec:implementation} discusses the molecular-lattice implementation and experimental signatures.
We close with broader implications and outlook in Sec.~\ref{sec:discussion}.

\section{Model and Numerical Approach}
\label{sec:model}

We consider hard-core bosons on a two-dimensional square lattice with Hamiltonian
\begin{equation}
H=
-t\sum_{\langle i,j\rangle}\left(b_i^\dagger b_j+\mathrm{H.c.}\right)
-\mu\sum_i n_i
+\sum_{i<j} V_{ij}\, n_i n_j,
\label{eq:H}
\end{equation}
where \(b_i^\dagger\) and \(b_i\) create and annihilate a boson at site \(i\), \(n_i=b_i^\dagger b_i\), \(t\) is the nearest-neighbor hopping amplitude, and \(\mu\) is the chemical potential.
The hard-core constraint restricts the onsite occupation to \(n_i=0,1\).
Throughout the paper we use \(t\) as the base energy scale.

\begin{figure*}[t]
\centering
\includegraphics[width=0.82\textwidth]{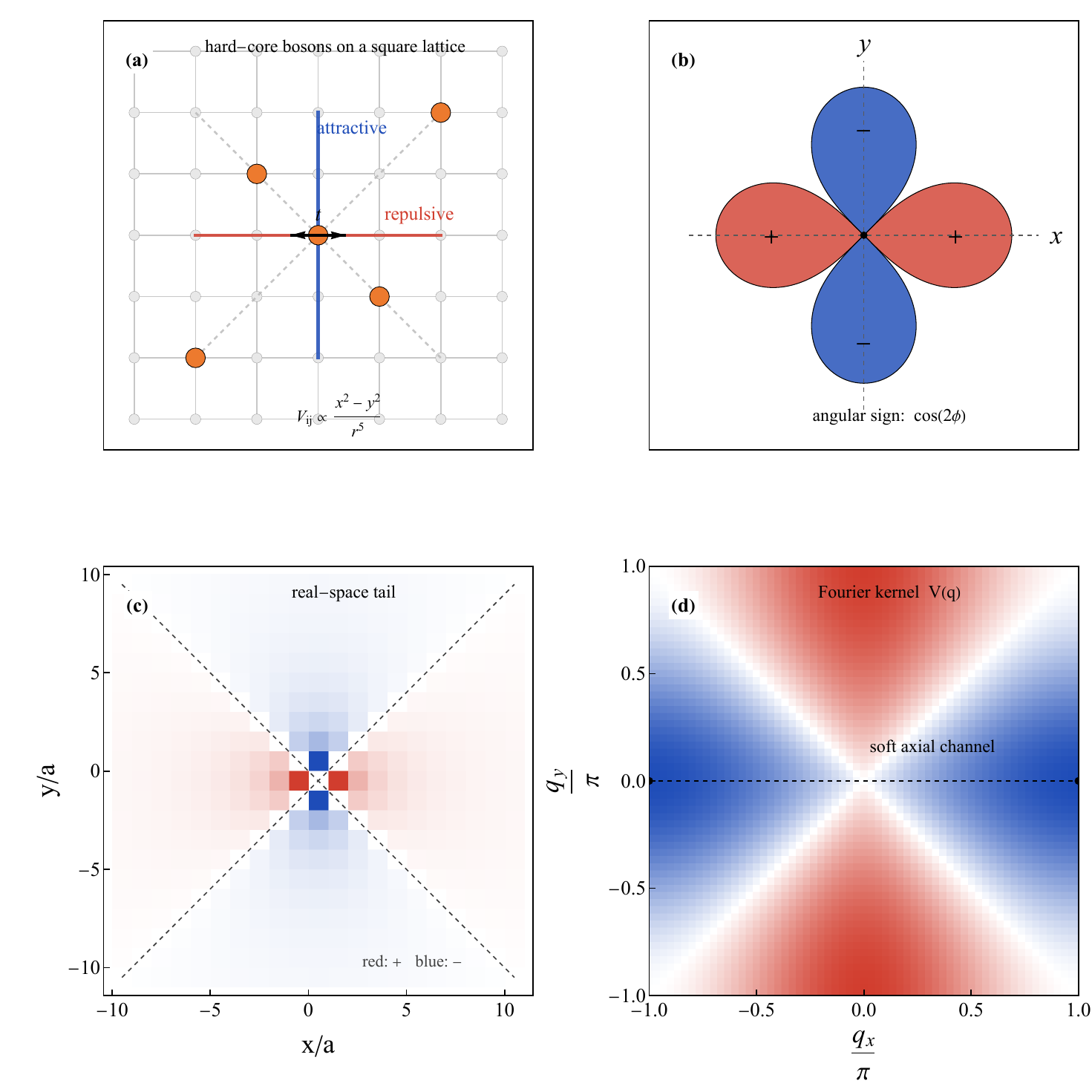}
\caption{
\textbf{Model and interaction geometry.}
(a) Hard-core bosons hop with amplitude \(t\) on a square lattice and interact through a long-range, sign-changing density interaction \(V_{ij}=\mathcal V(\mathbf r_{ij})\).
Red and blue links indicate the repulsive and attractive axial sectors, while the dashed diagonals mark the nodal directions.
(b) Angular part of the interaction, proportional to \(\cos 2\phi\), illustrating its \(d_{x^2-y^2}\)-like sign structure.
(c) Representative real-space interaction window computed directly from Eq.~\eqref{eq:Vxy}; a sign-preserving compressed color scale is used for visibility, with warm and cool colors denoting positive and negative matrix elements, respectively.
(d) Representative dense-grid momentum-space kernel \(\widetilde{\mathcal V}(\mathbf q)\).
The same color convention is used in momentum space, where the lowest region lies in the axial \(q_y=0\) channel rather than on the diagonal or checkerboard momenta, anticipating the stripe-family instability diagnosed below.
}
\label{fig:model}
\end{figure*}

The offsite matrix element is \(V_{ij}=\mathcal V(\mathbf r_{ij})\), with the sign-changing \(d_{x^2-y^2}\)-like lattice kernel
\begin{equation}
\begin{aligned}
\mathcal V(\mathbf r)
&=
V\,\frac{\cos(2\phi)}{r^3}
\\
&=
V\,\frac{x^2-y^2}{(x^2+y^2)^{5/2}},
\qquad
\mathbf r=(x,y)\neq (0,0).
\end{aligned}
\label{eq:Vxy}
\end{equation}
Here each lattice site \(i\) has coordinate \(\mathbf r_i=(x_i,y_i)\) in units of the lattice spacing \(a\), which is set to unity throughout.
For a pair of sites, \(\mathbf r_{ij}=\mathbf r_i-\mathbf r_j=(x,y)\) is the relative lattice displacement.
The variables \(r=\sqrt{x^2+y^2}\) and \(\phi\) are the polar coordinates of this in-plane relative displacement, and the second equality follows from \(\cos(2\phi)=(x^2-y^2)/r^2\).
We use \(V_{ij}\) for the pair matrix element, \(\mathcal V(\mathbf r)\) for the real-space lattice interaction, \(\widetilde{\mathcal V}(\mathbf q)\) for its lattice Fourier transform, and \(V\) without an argument for the scalar interaction scale.
Thus the \(x\) and \(y\) appearing in \(\mathcal V(\mathbf r)\) are relative coordinate components, not additional site labels.
For this choice, the interaction is repulsive along the \(x\) axis, attractive along the \(y\) axis, and vanishes along the nodal directions \(x=\pm y\).
Changing the sign of \(V\) reverses the sign of the same kernel, which is mathematically equivalent to interchanging the \(x\) and \(y\) directions; without loss of generality we focus on \(V>0\).
The essential point is that the interaction distinguishes lattice directions already at the two-body level, without introducing anisotropic hopping, tilted external fields, or nonequilibrium driving into the kinetic term.
Figure~\ref{fig:model} summarizes this lattice interaction geometry in real and momentum space.

Equation~\eqref{eq:Vxy} is the point-particle lattice version of a continuum in-plane anisotropic harmonic.
To keep the notation distinct, we denote this idealized long-distance continuum tail by
\begin{equation}
U_{D_2}^{\rm 2D}(R,\phi)
=
\frac{D_2}{R^3}\cos(2\phi),
\label{eq:continuumD2}
\end{equation}
where \(R\) is the physical continuum separation.
For lattice sites separated by \(R=a r\), a deep-lattice projection gives
\(U_{D_2}^{\rm 2D}(a r,\phi)=(D_2/a^3)\cos(2\phi)/r^3\).
In Eq.~\eqref{eq:Vxy}, the factor \(D_2/a^3\), together with small Wannier-smearing renormalizations, is absorbed into the single lattice energy scale \(V\).
Thus \(U_{D_2}^{\rm 2D}\) denotes only the asymptotic \(D_2\) harmonic, while \(U_{\rm eff}\), used later in Sec.~\ref{sec:implementation}, denotes the full microwave-dressed continuum potential including shielding structure, residual isotropic terms, and short-range physics.

The physical motivation comes from microwave-dressed polar molecules, where shielding fields create a strong short-range repulsive core while leaving a tunable long-range dipolar tail~\cite{GorshkovPRL2008,KarmanPRL2018,MatsudaScience2020,AndereggScience2021,SchindewolfNature2022,LinPRX2023,BigagliNatPhys2023,ZhangPRL2024Magic,StevensonPRL2024,KarmanPRXQ2025,DengNatCommun2025}.
Static-field and microwave-shielding analyses further show that elastic scattering and loss can be tuned over broad parameter regimes~\cite{MukherjeeHutsonPRR2024,DuttaPRR2025}.
Field-linked and microwave-dressed experiments demonstrate that the effective interaction can be probed and controlled through resonances, dressed-state spectroscopy, magic trapping, and few-body loss channels~\cite{ChenNature2023,ChenNature2024Tetramer,ZhangPRL2024Magic,StevensonPRL2024}.
Theory also suggests that additional polarization control can imprint a nontrivial in-plane anisotropy on the effective interaction~\cite{DengNatCommun2025,ZhangArxiv2025Supersolid}.
In an optical-lattice realization, the lattice depth controls the tunneling \(t\), while the dressed dipole strength and polarization geometry control the interaction scale \(V\) and its sign-changing angular form.
The mean filling is set by preparation and by the global chemical potential, or equivalently by the total molecule number in a fixed-\(N\) experiment; a weak residual trap does not set the particle number by itself, but produces a spatially varying local chemical potential and therefore a range of local fillings across the cloud.

We therefore use Eq.~\eqref{eq:Vxy} as a clean \(D_2\)-only lattice benchmark: the hard-core constraint represents the dominant short-range repulsion, while the offsite interaction retains the sign-changing \(1/r^3\) tail whose angular geometry we study.
A protocol-specific calculation would instead start from the full \(U_{\rm eff}(R,\phi)\) and perform the Wannier projection.
This deliberate limiting choice isolates a clean many-body question: how does a non-axisymmetric dipolar geometry alone reorganize the competition between superfluidity and density ordering on a square lattice?

We study Eq.~\eqref{eq:H} with path-integral quantum Monte Carlo simulations implemented with the worm algorithm in the grand-canonical ensemble~\cite{ProkofevPLA1998,ProkofevJETP1998}.
The simulations scan \(V/t\) and \(\mu/t\) on finite \(L\times L\) lattice with periodic boundary conditions; unless stated otherwise, the main figures use \(L=24\) and inverse temperature \(\beta=L\).
In the simulations, each pair interaction \(V_{ij}\) with \(i<j\) is evaluated from Eq.~\eqref{eq:Vxy} using the relative lattice displacement defined with periodic boundary conditions; the interaction range spans the full finite lattice used in the calculation.

The density and compressibility are
\begin{equation}
n=\frac{\avg{N}}{N_s},\qquad
\kappa=\frac{\beta}{N_s}\left(\avg{N^2}-\avg{N}^2\right),
\label{eq:density_kappa}
\end{equation}
Here \(N_s=L^2\) is the fixed number of lattice sites, \(N=\sum_i n_i\) is the instantaneous total particle number in a Monte Carlo configuration, and \(n=\avg{N}/N_s\) is the filling.
The directional superfluid response is measured from winding-number fluctuations~\cite{PollockPRB1987},
\begin{equation}
\rho_{s,x}=\frac{\avg{W_x^2}}{2t\beta},
\qquad
\rho_{s,y}=\frac{\avg{W_y^2}}{2t\beta},
\qquad
\rho_s=\frac{\rho_{s,x}+\rho_{s,y}}{2}.
\label{eq:rhosxy}
\end{equation}
We also use \(A_\rho=(\rho_{s,x}-\rho_{s,y})/(\rho_{s,x}+\rho_{s,y})\) to quantify the stiffness anisotropy.

Density order is diagnosed from the equal-time structure factor on the finite reciprocal grid \(\mathbf Q=(2\pi m_x/L,2\pi m_y/L)\).
We use the site-normalized convention
\begin{equation}
S(\mathbf Q)
=
\frac{1}{N_s^2}
\left\langle
\left|
\sum_j
n_j e^{i\mathbf Q\cdot\mathbf r_j}
\right|^2
\right\rangle .
\label{eq:Sq}
\end{equation}
For the first-order diagnostic in Fig.~\ref{fig:V218_first_order}, we also retain measurement-resolved stripe structure factors to distinguish stable coexistence from switching between sectors.
To compare order across fillings, momenta are folded componentwise into \(q_\alpha^{\rm fold}/\pi\in[0,1]\), and the directly sampled structure-factor signal is grouped into the axial and diagonal families
\begin{align}
S_x^\star
&=
\sum_{\substack{\mathbf q:\\
q_y^{\mathrm{fold}}/\pi=0,\,
q_x^{\mathrm{fold}}/\pi>0}}
S(\mathbf q),
\nonumber\\
S_y^\star
&=
\sum_{\substack{\mathbf q:\\
q_x^{\mathrm{fold}}/\pi=0,\,
q_y^{\mathrm{fold}}/\pi>0}}
S(\mathbf q),
\nonumber\\
S_d^\star
&=
\sum_{\substack{\mathbf q:\\
q_x^{\mathrm{fold}}/\pi=q_y^{\mathrm{fold}}/\pi>0}}
S(\mathbf q).
\label{eq:familysum}
\end{align}

\begin{figure*}[t]
\centering
\includegraphics[width=\textwidth]{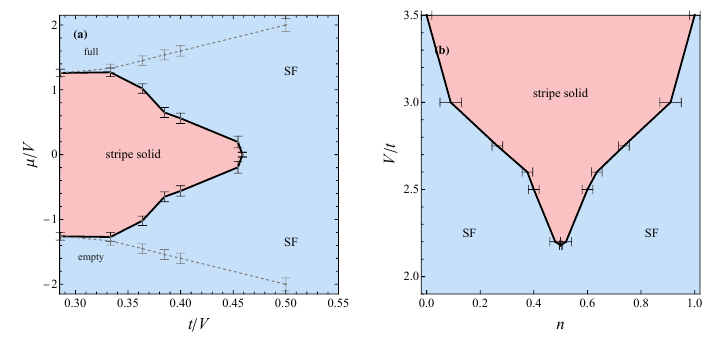}
\caption{
\textbf{Ground-state phase diagram.}
(a) Ground state phase diagram in the \((t/V,\mu/V)\) plane.
Blue and pink denote the superfluid and stripe-solid regimes, respectively; the solid black boundary is bracketed from the combined compressibility, superfluid-stiffness, and stripe-family diagnostics.
Dashed curves indicate the crossover toward the trivial empty and full limits.
(b) The same boundary replotted in the \((n,V/t)\) plane, highlighting the particle-hole-related structure about half filling.
The stripe lobe first appears near \(n=1/2\), broadens with increasing \(V/t\), and closes near \(V/t=2.18\), as analyzed in Fig.~\ref{fig:V218_first_order}.
}
\label{fig:phase}
\end{figure*}

The phases in Fig.~\ref{fig:phase} below are based on these diagnostics: superfluid regions have finite winding response, finite compressibility, and no dominant stripe-family structure-factor signal, while stripe-solid regions show suppressed \(\rho_s\), suppressed \(\kappa\), and a strong signal in the selected axial family.
The phase boundaries are placed between neighboring parameter points where the combined diagnostics change from superfluid stiffness to structure factor of stripe-solid.

\section{Physical Picture}
\label{sec:analytic}

The numerical results below are organized by two simple pieces of theory.
First, the hard-core constraint permits the exact spin-\(1/2\) identification
\begin{equation}
b_i^\dagger \rightarrow S_i^+,\qquad
b_i \rightarrow S_i^-,\qquad
n_i=S_i^z+\frac{1}{2},
\label{eq:spinmap}
\end{equation}
which maps the superfluid to in-plane spin coherence and density order to modulation of \(S^z\).
The Hamiltonian becomes
\begin{align}
H_{\mathrm{spin}}
=&
-2t\sum_{\langle i,j\rangle}
\left(S_i^xS_j^x+S_i^yS_j^y\right)
\nonumber\\
&+\sum_{i<j}V_{ij}S_i^zS_j^z
-h\sum_i S_i^z
+\mathrm{const.}
\label{eq:Hspin}
\end{align}
where the longitudinal field contains the Hartree shift,
\begin{equation}
h=\mu-\frac{V_H}{2},
\qquad
V_H=\sum_{j(\neq i)}V_{ij},
\label{eq:hartree_shift}
\end{equation}
for translationally invariant pair couplings.
The problem is therefore a competition between short-range \(XY\) coherence from hopping and a long-range sign-changing Ising interaction fixed by Eq.~\eqref{eq:Vxy}.

Second, the ordering channel is exposed by the lattice Fourier transform
\begin{equation}
\widetilde{\mathcal V}(\mathbf q)=
\sum_{\mathbf r\neq 0}
\mathcal V(\mathbf r)e^{i\mathbf q\cdot\mathbf r}.
\label{eq:Vq}
\end{equation}
with \(\widetilde{\mathcal V}(\mathbf 0)=V_H\).
A hard-core Gutzwiller expansion around a uniform superfluid with density \(n\) gives the quadratic density kernel
using the convention
\(\delta n_i=N_s^{-1/2}\sum_{\mathbf q}\delta n_{\mathbf q}e^{i\mathbf q\cdot\mathbf r_i}\) and
\(\delta n_{\mathbf q}=N_s^{-1/2}\sum_i\delta n_i e^{-i\mathbf q\cdot\mathbf r_i}\):
\begin{align}
\delta E_G^{(2)}
&=
\frac{1}{2}
\sum_{\mathbf q}
\mathcal K(\mathbf q,n)
\left|\delta n_{\mathbf q}\right|^2,
\nonumber\\
\mathcal K(\mathbf q,n)
&=
\mathcal K_t(\mathbf q,n)+\widetilde{\mathcal V}(\mathbf q),
\label{eq:gutz_kernel}
\end{align}
where the square-lattice hopping contribution is
\begin{equation}
\mathcal K_t(\mathbf q,n)
=
\frac{2t}{n(1-n)}
\left[
1-\frac{(1-2n)^2}{2}
\left(\cos q_x+\cos q_y\right)
\right].
\label{eq:gutz_kinetic_kernel}
\end{equation}
The kinetic term is positive and respects the square-lattice symmetry, whereas the interaction kernel carries the \(d_{x^2-y^2}\) antisymmetry of the real-space tail.
Because \(\mathcal V(y,x)=-\mathcal V(x,y)\), the Fourier transform obeys \(\widetilde{\mathcal V}(q_y,q_x)=-\widetilde{\mathcal V}(q_x,q_y)\) for finite-size pair couplings that preserve this \(x\leftrightarrow y\) pairing.
Already at the nearest-neighbor level, \(\widetilde{\mathcal V}_{\rm nn}(\mathbf q)=2V\cos q_x-2V\cos q_y\), so the \((q,0)\) and \((0,q)\) axial channels have opposite signs.
The kernel \(\mathcal K(\mathbf q,n)\) is the quadratic energy cost of adding a small density modulation at wave vector \(\mathbf q\) to the uniform state.
The modulation with the smallest \(\mathcal K\) is therefore the one that becomes unstable first within this mean-field picture.
For the present interaction, this minimum lies in the axial \((q,0)\) family, explaining why the ordered state selects stripes running parallel to \(y\).
The value of \(q\) can still change with filling because \(\mathcal K_t\) depends on \(n\).
We use this kernel mainly to explain the selected family; deep in the stripe solid, the precise period is controlled by the hard-core packing of particles on the lattice rather than by the quadratic expansion alone.

The Hartree shift fixes the particle-hole convention.
For hard-core bosons, the transformation \(b_i\leftrightarrow b_i^\dagger\), \(n_i\rightarrow1-n_i\) maps \(H(\mu)\) to \(H(V_H-\mu)\) up to a constant, so half filling occurs at \(\mu_{1/2}=V_H/2\) and the density obeys
\begin{equation}
n(\mu_{1/2}+\delta\mu)
=
1-n(\mu_{1/2}-\delta\mu).
\label{eq:ph_density}
\end{equation}
For the full symmetric displacement set used in the simulations, the pairwise \(x\leftrightarrow y\) cancellation gives \(V_H=0\) within floating-point roundoff; hence \(\mu_{1/2}=0\).
This symmetry explains the near-reflection symmetry of the phase diagram about half filling.
The same interaction anisotropy also explains why the larger-\(t/V\) superfluid can already have \(\rho_{s,x}\neq\rho_{s,y}\): although the hopping is isotropic, the sign-changing interaction produces different density correlations along \(x\) and \(y\), leading to different winding-number fluctuations in the two directions.

\section{Results}
\label{sec:results}

The physical picture above leads to two concrete expectations.
First, the density instability should be selected in an axial stripe family rather than in a checkerboard or diagonal channel.
Second, the superfluid response should reflect the explicit \(C_4\)-breaking anisotropy of the interaction even before static density order appears.
The numerical results below confirm both expectations and show how the selected stripe period changes with filling.

\subsection{Ground-state phase diagram}
\label{subsec:phase_diagram}

The resulting ground-state phase diagram is shown in Fig.~\ref{fig:phase}.
Panel (a) keeps the chemical potential as the control variable used in the simulations, while panel (b) reorganizes the same boundary estimates by filling.
This second representation is useful because the hard-core particle-hole symmetry discussed in Sec.~\ref{sec:analytic} acts directly as \(n\rightarrow 1-n\).
Because the finite-size interaction preserves the \(x\leftrightarrow y\) cancellation of the sign-changing tail, the Hartree shift vanishes, \(V_H=0\).
The particle-hole symmetry discussed in Sec.~\ref{sec:analytic} then places half filling at \(\mu_{1/2}=0\), explaining why the phase diagram is centered around \(\mu=0\).

The map shows a simple competition.
At larger \(t/V\), or equivalently smaller \(V/t\), kinetic delocalization dominates and the compressible region is superfluid.
Equivalently, increasing \(V/t\) first stabilizes stripe order near half filling, where the density modulation gains the most interaction energy.
With further increasing \(V/t\), the stripe-solid region broadens in filling and extends toward both the particle- and hole-doped sides, while remaining centered near \(n=1/2\).
The approximate symmetry of Fig.~\ref{fig:phase}(b) about half filling reflects the hard-core particle-hole structure of the Hamiltonian.

\begin{figure}[t]
\centering
\includegraphics[width=\columnwidth]{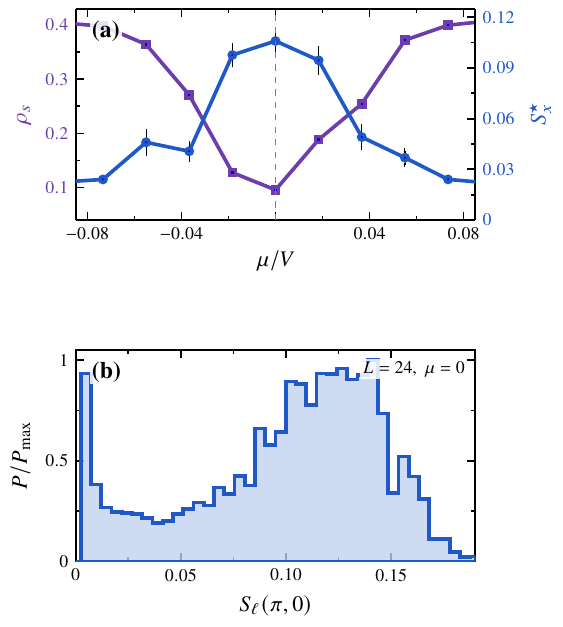}
\caption{
\textbf{First-order closure of the stripe-solid lobe at \(t/V\simeq0.459\).}
(a) \(L=24\) parameter sequence near half filling: the left axis gives the superfluid stiffness \(\rho_s\), while the right axis gives the stripe-family structure factor \(S_x^\star\).
(b) Measurement-resolved normalized distribution of the fixed stripe structure factor \(S_\ell(\pi,0)\) at \(\mu=0\); the horizontal axis is the unscaled \(S_\ell(\pi,0)\) value.
The simultaneous dip in the averaged winding response, growth of the stripe structure factor, and two-sector structure of this distribution indicate first-order switching near the lobe closure rather than a separate stable supersolid phase.
}
\label{fig:V218_first_order}
\end{figure}

\begin{figure}[t]
\centering
\includegraphics[width=\columnwidth]{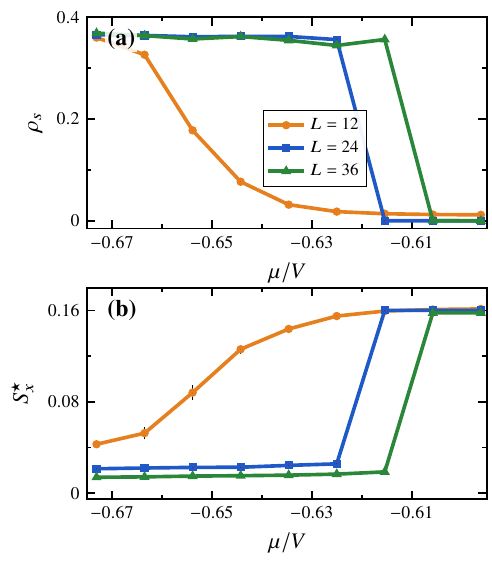}
\caption{\textbf{System-size dependence of a representative stripe-solid boundary.}
Representative transition region at \(t/V\simeq0.385\) for \(L=12,24,36\), plotted as a function of \(\mu/V\).
(a) Superfluid stiffness \(\rho_s\); (b) stripe-family structure factor \(S_x^\star\).
The \(L=12\) response is broadened over several points, whereas the \(L=24\) and \(L=36\) curves show a much sharper jump, consistent with first-order behavior.
Error bars denote statistical uncertainties; the drift of the jump position with \(L\) gives the finite-size uncertainty used when assigning the phase boundary in Fig.~\ref{fig:phase}.
}
\label{fig:finiteL_first_order}
\end{figure}

Figure~\ref{fig:V218_first_order} illustrates the evidence used to identify the closure of the stripe lobe.
At \(t/V\simeq0.459\), averaged observables near half filling show the potentially misleading combination of an enhanced stripe-family structure-factor signal and only partially suppressed winding response.
Taken alone, such averaged quantities could be read as a narrow coexistence window between stripe order and superfluid response.
The measurement-resolved fixed-stripe structure factor shows a different structure.
The normalized distribution of \(S_\ell(\pi,0)\) separates low- and high-\(S_\ell(\pi,0)\) sectors instead of forming a single narrow peak around one homogeneous state.
This is the coexistence signature expected near a first-order boundary: within the narrow closure region, the simulation samples superfluid-like sectors with residual winding response and stripe-solid-like sectors with a large stripe structure factor, rather than one homogeneous phase with simultaneous long-range coherence and density order.
We therefore use this \(t/V\simeq0.459\) parameter sequence to close the stripe lobe between the superfluid and stripe-solid regimes, rather than to identify a stable supersolid phase.
A complementary comparison across system sizes is performed on a representative stripe-solid boundary, \(t/V\simeq0.385\), rather than exactly at the lobe closure, as shown in Fig.~\ref{fig:finiteL_first_order}.
For the smallest size the change is visibly broadened, while for \(L=24\) and \(L=36\) the drop of \(\rho_s\) and the rise of \(S_x^\star\) become much more abrupt, supporting the same first-order interpretation of the stripe boundary.
For a molecular quantum simulator, this distinction is practical: shot- or bin-averaged measurements can mimic coexistence even when the underlying samples switch between competing sectors.

\subsection{\texorpdfstring{Larger-\(t/V\) regime: anisotropic superfluid at \(t/V=0.5\)}{Larger-t/V regime: anisotropic superfluid at t/V=0.5}}
\label{subsec:V2_SF}

\begin{figure}[tbp]
\centering
\includegraphics[width=\columnwidth]{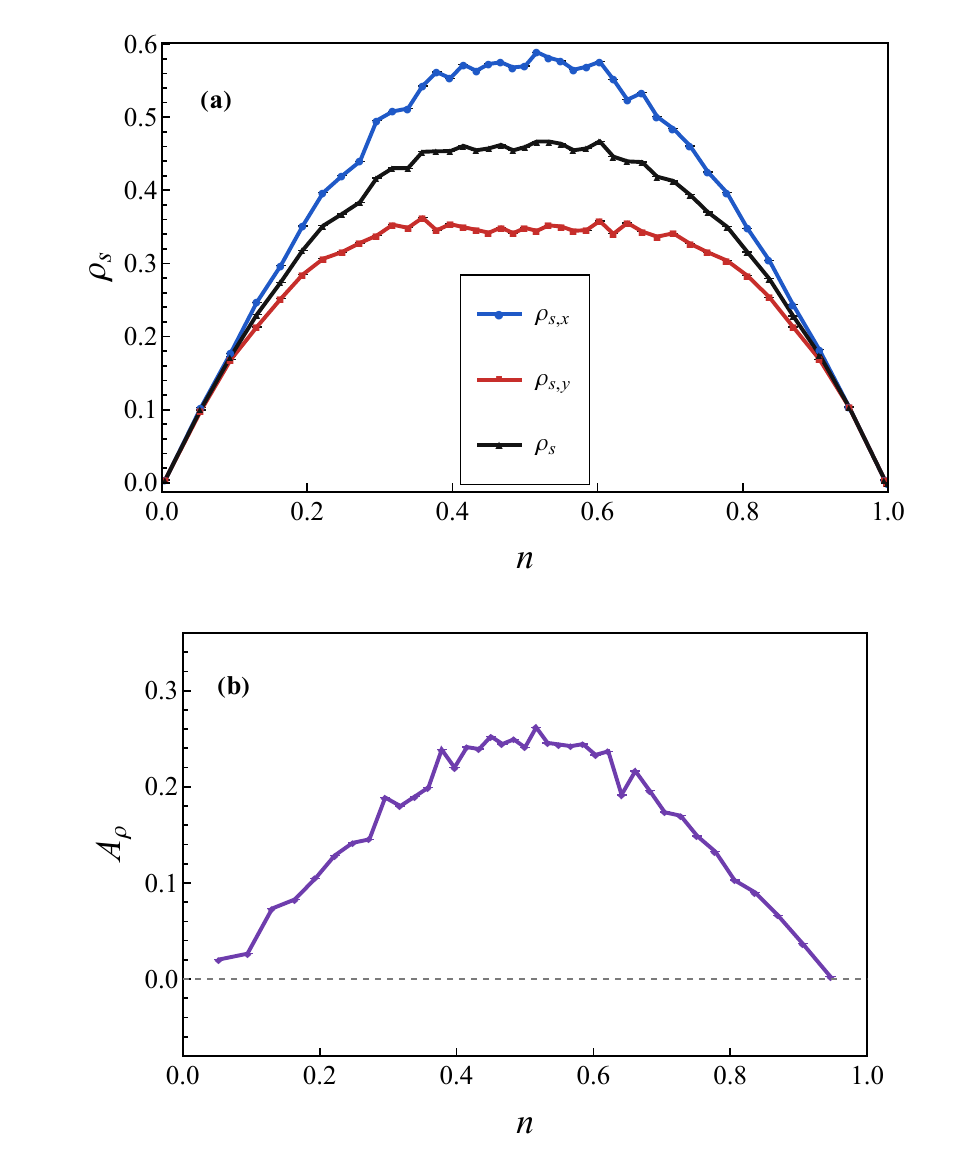}
\caption{
\textbf{Anisotropic superfluid response at larger hopping ratio, \(t/V=0.5\).}
Representative \(t/V=0.5\) case for \(L=24\), plotted against filling \(n\), where hopping is relatively stronger than in the stripe-ordered regime and the system remains superfluid throughout the compressible window.
(a) Directional superfluid stiffnesses \(\rho_{s,x}\), \(\rho_{s,y}\), and their average \(\rho_s\).
The superfluid response is finite over a broad interval and vanishes only near the trivial empty and full limits.
A clear directional imbalance, \(\rho_{s,x}>\rho_{s,y}\), persists across most of the superfluid regime.
(b) Stiffness anisotropy factor \(A_\rho\).
It remains finite inside the superfluid window and decreases toward zero near the density boundaries.
Thus, even before stripe order forms, the interaction already imprints a pronounced directional preference on the superfluid response.
}
\label{fig:V2_SF}
\end{figure}

We first examine the representative larger-hopping case \(t/V=0.5\), shown in Fig.~\ref{fig:V2_SF}(a).
In this regime the system remains superfluid throughout the compressible window: both \(\rho_{s,x}\) and \(\rho_{s,y}\) are finite except near the trivial empty and full limits.
Nevertheless, the superfluid is not isotropic.
Over most of the accessible filling range, the stiffness along the \(x\) direction exceeds that along \(y\), so the same interaction geometry that ultimately produces stripe order already manifests itself as anisotropic phase coherence at larger \(t/V\).
Figure~\ref{fig:V2_SF}(b) summarizes the same directional imbalance through the anisotropy factor \(A_\rho\).

The superfluid and stripe-ordered regimes are therefore two manifestations of the same interaction geometry: at larger \(t/V\) it distorts the phase stiffness, while at smaller \(t/V\) it overcomes kinetic delocalization and locks the system into directional density order.

\subsection{\texorpdfstring{Stripe-ordered regime and directional selection at \(t/V=1/3\)}{Stripe-ordered regime and directional selection at t/V=1/3}}
\label{subsec:V3_family}

\begin{figure}[tbp]
\centering
\includegraphics[width=\columnwidth]{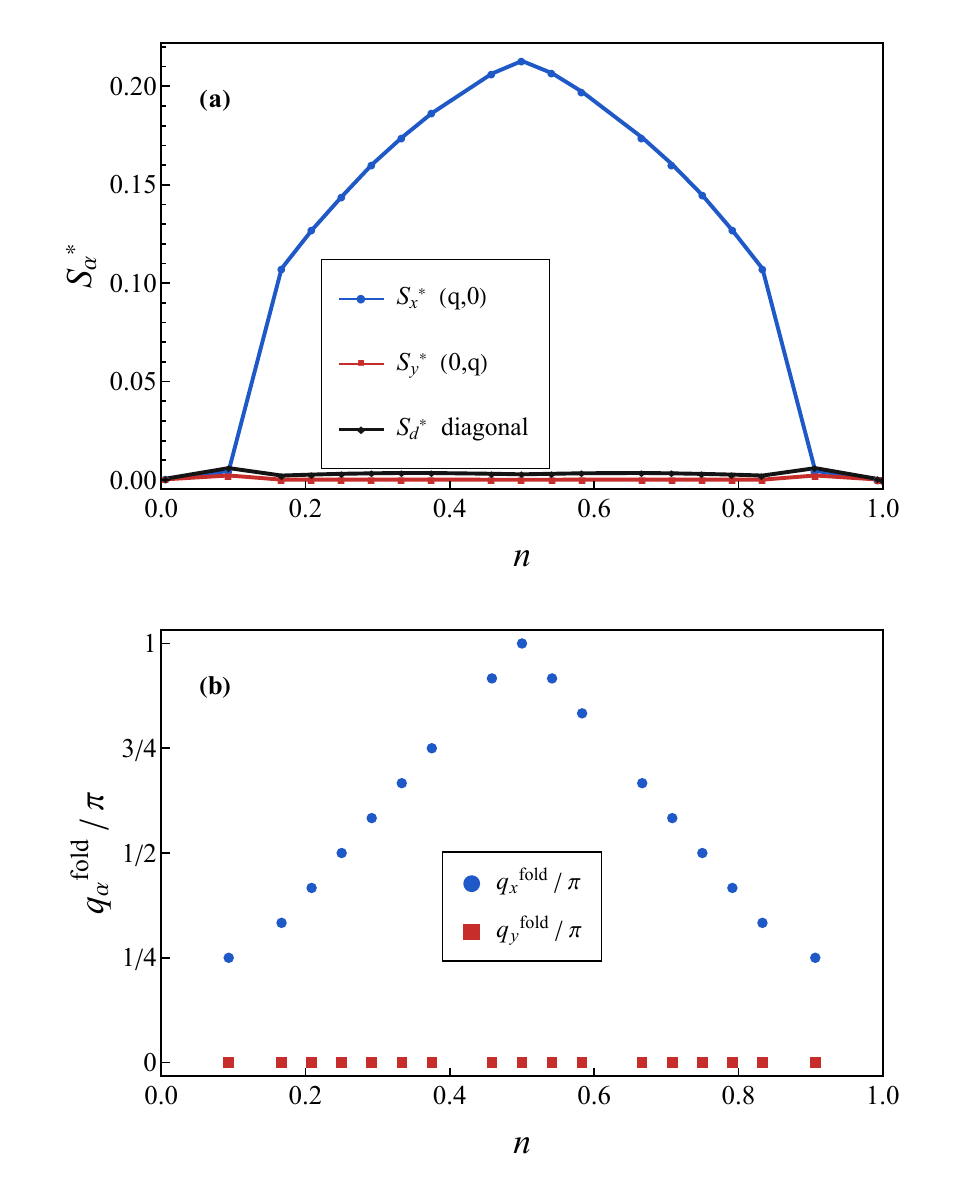}
\caption{\textbf{Stripe-family selection and folded ordering vectors at \(t/V=1/3\).}
(a) Directly accumulated family-resolved structure factors \(S_\alpha^\star\) for \(L=24\) as functions of filling \(n\), using the normalization in Eq.~\eqref{eq:Sq}.
The \((q,0)\) axial family develops a large signal over a broad intermediate-filling window; in real space this corresponds to \(y\)-oriented stripes, because the density modulation is along \(x\).
The competing \((0,q)\) axial family and the diagonal \((q,q)\) family remain strongly suppressed.
(b) Leading folded ordering vector extracted from the strongest inequivalent folded \(S(\mathbf q)\) peak at each filling, after merging symmetry-related folded momenta.
The leading peak remains in the \((q,0)\) sector while \(q_x^{\mathrm{fold}}/\pi\) evolves with filling, showing period selection inside the selected stripe family.
}
\label{fig:V3_family}
\end{figure}

The smaller-hopping case \(t/V=1/3\) enters a qualitatively different regime, as shown by the family-resolved structure factors in Fig.~\ref{fig:V3_family}(a).
This panel plots the directly accumulated \(S_\alpha^\star\) defined in Eq.~\eqref{eq:familysum}.
The filling \(n\) is obtained from the average particle number.
Over the sampled intermediate-filling window, the \((q,0)\) axial family acquires a large structure-factor signal, while the \((0,q)\) and \((q,q)\) families remain strongly suppressed.
Because an ordering wave vector \((q,0)\) modulates the density along \(x\), this corresponds in real space to stripes running parallel to \(y\).
This identifies the ordered regime as a stripe-dominated state selected into the \((q,0)\) directional channel by the interaction geometry.

Figure~\ref{fig:V3_family}(b) tests whether the stripe solid is characterized by one fixed ordering wave vector.
For each filling, we identify the largest folded peak of the directly measured \(S(\mathbf q)\), without imposing a family constraint.
The peak always lies in the \((q,0)\) sector, but its \(q_x^{\mathrm{fold}}/\pi\) value changes with \(n\).
Thus the robust feature is the selected stripe direction, not a single commensurate wave vector such as \((\pi,0)\).
The Gutzwiller analysis below is therefore used to explain the family selection, while the detailed filling-dependent period reflects nonlinear hard-core packing within the ordered state.

\begin{figure}[tbp]
\centering
\includegraphics[width=\columnwidth]{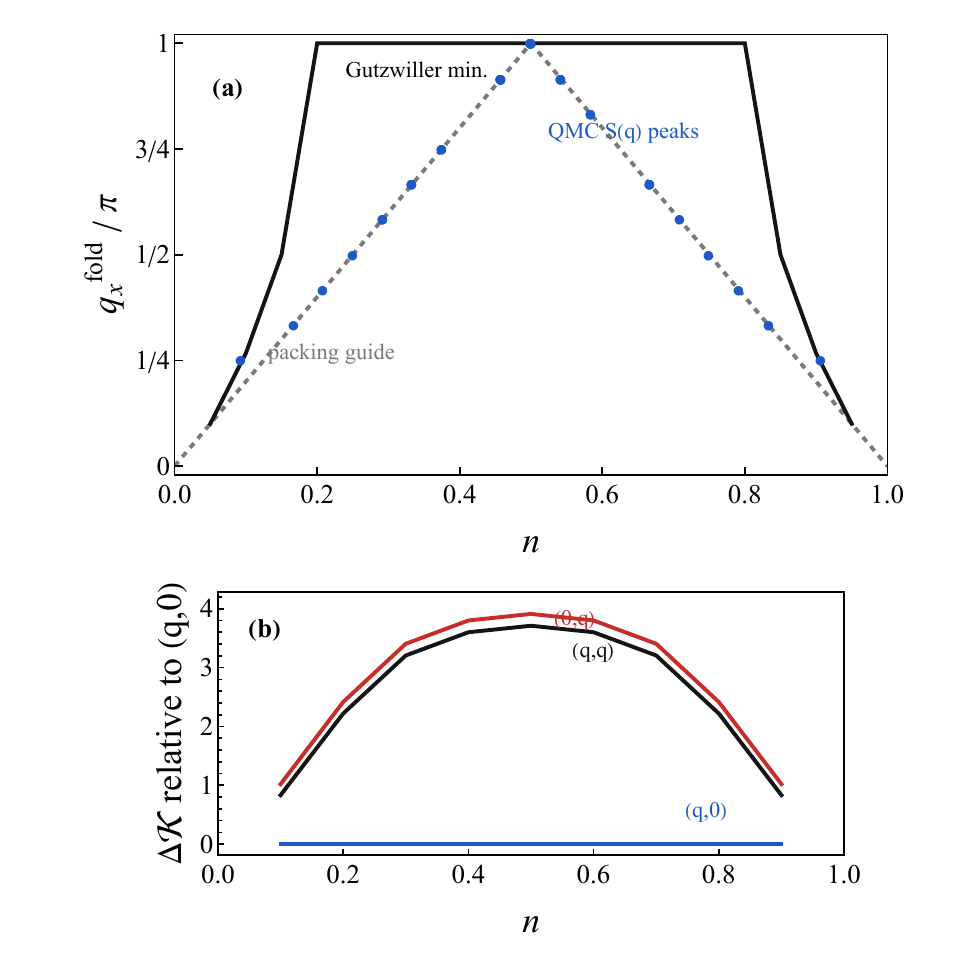}
\caption{
\textbf{Mean-field soft-mode diagnostic and measured stripe period.}
(a) Comparison between the leading folded \(S(\mathbf q)\) peak from the \(L=24\) ordered window of Fig.~\ref{fig:V3_family}(b), a simple stripe-packing guide \(q_x^{\mathrm{fold}}/\pi=2\min(n,1-n)\), and the minimum of the hard-core Gutzwiller kernel \(\mathcal K(\mathbf q,n)=\mathcal K_t(\mathbf q,n)+\widetilde{\mathcal V}(\mathbf q)\) at \(t/V=1/3\).
The Gutzwiller minimum is evaluated on a symmetric \(L=60\) reciprocal grid and lies in the \(q_y^{\mathrm{fold}}/\pi=0\) sector over the plotted range.
(b) Relative family softening \(\Delta\mathcal K_{\rm fam}=\min_{\rm fam}\mathcal K-\min_{(q,0)}\mathcal K\), confirming that the \((0,q)\) and \((q,q)\) channels remain harder than the selected \((q,0)\) family.
The measured structure-factor peak follows the same family but drifts close to the stripe-packing guide, indicating that the quadratic theory accounts for family selection while the detailed period is controlled by hard-core packing.
}
\label{fig:gutz_diagnostic}
\end{figure}

Figure~\ref{fig:gutz_diagnostic} provides the corresponding theory diagnostic at \(t/V=1/3\).
The hard-core Gutzwiller kernel captures the linear instability of the uniform state and identifies the axial \((q,0)\) family as the soft channel.
It is not a quantitative fit to the QMC ordering wave vector at every filling.
Instead, the comparison shows a division of roles: the interaction kernel explains why the system chooses an axial stripe family, while the hard-core density constraint explains why the ordering period reorganizes approximately with \(2\min(n,1-n)\).

\subsection{\texorpdfstring{Real-space evolution of stripe patterns at \(t/V=1/3\)}{Real-space evolution of stripe patterns at t/V=1/3}}
\label{subsec:V3_maps}

\begin{figure}[tbp]
\centering
\includegraphics[width=\columnwidth]{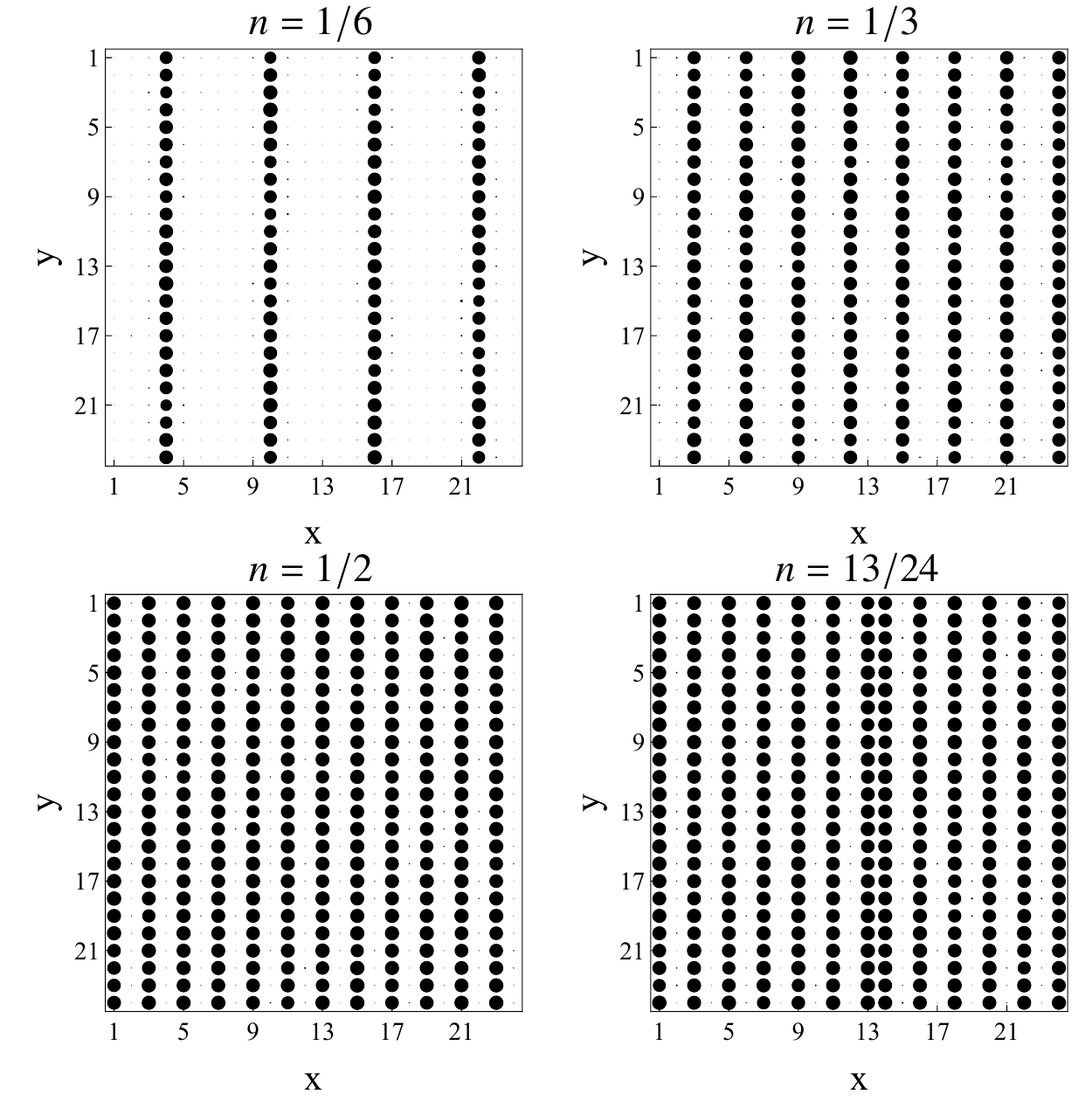}
\caption{
\textbf{Representative real-space stripe patterns at \(t/V=1/3\).}
Imaginary-time-averaged site-density maps on \(L=24\) clusters for four representative fillings at \(t/V=1/3\): \(n=1/6\), \(n=1/3\), \(n=1/2\), and \(n=13/24\).
In all cases, the particles form stripes running parallel to the \(y\) direction, corresponding to ordering wave vectors in the \((q,0)\) family.
As the filling increases, the stripe spacing decreases and the occupied regions broaden, with no indication of checkerboard or diagonal patterns.
}
\label{fig:V3_maps}
\end{figure}

The directional selection diagnosed above is directly reflected in the imaginary-time-averaged real-space density maps at \(t/V=1/3\) shown in Fig.~\ref{fig:V3_maps}.
For all representative fillings, the particles organize into stripes running parallel to the \(y\) direction, exactly as expected for ordering wave vectors in the \((q,0)\) family.
The filling evolution gives a direct real-space interpretation of Fig.~\ref{fig:V3_family}(b): the orientation is locked by the interaction geometry, while the stripe spacing and width evolve with density.
The ordered regime should therefore be viewed not as a sequence of unrelated commensurate crystals, but as one interaction-selected stripe family whose period changes with filling.

\section{Molecular-Lattice Implementation and Experimental Signatures}
\label{sec:implementation}

\subsection{Mapping to a molecular lattice simulator}

The minimal model in Eq.~\eqref{eq:H} can be obtained as the lowest-band lattice projection of a dressed molecular gas.
Here \(U_{\rm eff}\) denotes the full continuum interaction of the dressed molecules, not only the \(D_2\) harmonic \(U_{D_2}^{\rm 2D}\) retained in the minimal model.
For Wannier orbitals \(w_i(\mathbf r)=w(\mathbf r-\mathbf R_i)\), the lattice couplings are
\begin{align}
V_{ij}
&=
\int d\mathbf r\,d\mathbf r'\,
|w_i(\mathbf r)|^2 |w_j(\mathbf r')|^2
U_{\rm eff}(\mathbf r-\mathbf r'),
\label{eq:wannierV}
\\
U
&=
\int d\mathbf r\,d\mathbf r'\,
|w_i(\mathbf r)|^2 |w_i(\mathbf r')|^2
U_{\rm eff}(\mathbf r-\mathbf r'),
\label{eq:wannierU}
\end{align}
while the hopping \(t\) is controlled primarily by the lattice depth.
In the deep-lattice limit, where the Wannier orbitals are narrow compared with the lattice spacing, Eq.~\eqref{eq:wannierV} reduces to \(V_{ij}\simeq U_{\rm eff}(\mathbf R_i-\mathbf R_j)\) for \(i\neq j\).
The hard-core model studied here is the \(U/t\rightarrow\infty\) limit of this lattice projection.
For an explicitly engineered non-axisymmetric dressing, one can write the long-distance part of the dressed interaction schematically as
\begin{equation}
U_{\rm eff}(R,\phi)
=
\frac{D_0}{R^3}
+
\frac{D_2}{R^3}\cos 2\phi
+
U_{\rm short}(R,\phi)
+
\cdots ,
\label{eq:UeffExpansion}
\end{equation}
where \(D_0\) is the axisymmetric dipolar component, \(D_2\) is the engineered \(d_{x^2-y^2}\)-like harmonic, and \(U_{\rm short}\) contains short-range shielding and molecular-core physics.
This decomposition captures the long-range harmonic content of the dual-microwave potentials used in recent molecular-gas work.
For example, the non-axisymmetric NaCs droplet experiment writes the long-range microwave-dressed interaction as~\cite{ZhangNature2026}
\begin{align}
U_{\rm dd}(\mathbf R)
&=
\frac{\hbar^2}{m R^3}
\Big[
a_{d0}\left(1-3\cos^2\theta\right)
\nonumber\\
&\quad
+\sqrt{3}a_{d1}\cos\phi\sin 2\theta
\nonumber\\
&\quad
+\sqrt{3}a_{d2}\cos 2\phi\sin^2\theta
\Big],
\label{eq:UddLengths}
\end{align}
with \(a_{d2}\) controlled by the ellipticity of the microwave dressing field.
In the lattice plane, \(\theta=\pi/2\), the \(a_{d1}\) term drops out and Eq.~\eqref{eq:UddLengths} reduces to
\begin{equation}
U_{\rm dd}^{xy}(R,\phi)
=
\frac{\hbar^2}{m R^3}
\left(a_{d0}+\sqrt{3}a_{d2}\cos2\phi\right).
\label{eq:UddPlane}
\end{equation}
Thus \(D_0=\hbar^2a_{d0}/m\) and \(D_2=\sqrt{3}\hbar^2a_{d2}/m\).
Equivalently, in the notation of analytic dual-microwave effective potentials~\cite{DengNatCommun2025,ZhangArxiv2025Supersolid}, the long-range continuum potential is written as \(U_{\rm eff}(R,\theta,\phi)=R^{-3}[C_{3,0}(3\cos^2\theta-1)+C_{3,2}\sin^2\theta\cos2\phi]+\cdots\), which maps in the \(xy\) plane to \(D_0=-C_{3,0}\) and \(D_2=C_{3,2}\).
The numerical scale used below is tied to the concrete dressing parameters reported in the NaCs droplet experiment of Ref.~\cite{ZhangNature2026}.
That experiment used two microwave components: a linearly polarized \(\pi\) field along \(z\), coupling mainly to the \(|1,0\rangle\) rotational state, and an elliptically polarized \(\sigma\) field rotating in the \(xy\) plane, coupling to the \(|1,\pm1\rangle\) manifold.
The reported Rabi frequencies are \(\Omega_\sigma=2\pi\times8.0(1)\,{\rm MHz}\) and \(\Omega_\pi=2\pi\times7.1(1)\,{\rm MHz}\), with detuning magnitudes \(|\Delta_\sigma|=2\pi\times8.0\,{\rm MHz}\) and \(|\Delta_\pi|=2\pi\times10.75\,{\rm MHz}\) in the convention of that experiment~\cite{ZhangNature2026}.
These microscopic microwave parameters are not used directly in the lattice calculation; they enter only through the long-range coefficient \(a_{d2}\) of the engineered anisotropic harmonic.
At the compensated point, where the axisymmetric coefficient is tuned to \(a_{d0}=0\), Ref.~\cite{ZhangNature2026} reports
\begin{equation}
a_{d2}(\xi)
=
\left(5.31\times10^4a_0\right)\sin(2\xi),
\label{eq:ad2Nature}
\end{equation}
where \(\xi\) is the ellipticity angle of the \(\sigma\)-polarized field.
It also estimates residual terms from microwave-field imperfections of order \(|a_{d0}|<500a_0\), \(|a_{d1}|<1200a_0\), and \(|a_{d2}|<600a_0\) near compensation~\cite{ZhangNature2026}.
The present hard-core benchmark retains only the projected long-range \(D_2\) component in \(V_{ij}\), treats the short-range part as a large onsite exclusion, and leaves finite residual \(D_0\) corrections to the Wannier-projection upgrade described by Eq.~\eqref{eq:wannierV}.
The anisotropic coefficient can be expressed as
\begin{equation}
D_2=\sqrt{3}\frac{\hbar^2 a_{d2}}{m},
\label{eq:D2ad2}
\end{equation}
where \(a_{d2}\) is the effective anisotropic dipolar length.
For the lattice estimates we choose \(a_{d2}^{\rm ref}=1.3\times10^4a_0\) as a convenient reference anisotropic length.
Using Eq.~\eqref{eq:ad2Nature}, this corresponds to \(\xi\simeq7^\circ\), within the strongly anisotropic regime explored in Ref.~\cite{ZhangNature2026}.
The scale conversion used in Table~\ref{tab:feasibility} is
\begin{align}
\frac{D_2^{\rm ref}}{h a^3}
&=
\frac{\sqrt{3}\hbar^2 a_{d2}^{\rm ref}}{m h a^3}
\simeq
510\,{\rm Hz},
\nonumber\\
\frac{V}{h}
&\simeq
\eta\frac{D_2^{\rm ref}}{h a^3},
&
\frac{V}{t}
&=
\frac{V/h}{t/h},
\label{eq:latticeScaleConversion}
\end{align}
where \(a=532\,{\rm nm}\), \(m=156.9\,{\rm amu}\), and \(\eta=a_{d2}/a_{d2}^{\rm ref}\).
Thus \(\eta\) is only a dimensionless knob specifying what fraction of the reference anisotropic dressing strength is used.
This is the point-particle, deep-lattice estimate of the nearest-neighbor anisotropic scale; a full Wannier projection would replace the middle relation by Eq.~\eqref{eq:wannierV}.
The size of the omitted finite-band correction can be estimated in the harmonic approximation to the lowest-band Wannier orbital.
For a lattice depth \(sE_R\), where \(s\) is the dimensionless optical-lattice depth, the relative-coordinate Gaussian width is \(\ell_{\rm rel}/a\simeq 1/(\pi s^{1/4})\), giving \(\ell_{\rm rel}/a\simeq0.16\)--\(0.19\) for the depths in Table~\ref{tab:feasibility}.
For the offsite nearest-neighbor matrix element, where the relative-coordinate Gaussian is centered at \(R=a\), a harmonic Gaussian estimate suggests that Wannier smearing renormalizes the far-field \(D_2\cos2\phi/R^3\) tail only at the \(\sim10\%\) level for these depths.
We therefore keep the transparent point-particle values in Table~\ref{tab:feasibility} and regard Wannier smearing as a roughly ten-percent renormalization of the offsite couplings.
The onsite interaction and the exponentially small overlap-core contribution cannot be obtained from the asymptotic \(1/r^3\) tail alone; they require the full shielded dressed potential.
The relevant stripe-lobe window in Fig.~\ref{fig:phase} therefore corresponds to using only a tunable fraction of this reported anisotropic coefficient, or equivalently to choosing a modest effective dressed dipole moment in the simple \(V/h\) estimate below.
For the \(\eta\) values in Table~\ref{tab:feasibility}, Eq.~\eqref{eq:ad2Nature} would correspond to ellipticities of order \(0.5^\circ\)--\(1.3^\circ\), illustrating that the required lattice scale is below the large-ellipticity droplet regime and should be understood as the minimal \(D_2\)-only benchmark scale.

A useful estimate comes from a sinusoidal square lattice of depth \(sE_R\), lattice spacing \(a\), and molecular mass \(m\), for which
\begin{equation}
\frac{t}{E_R}
\simeq
\frac{4}{\sqrt{\pi}}s^{3/4}\exp[-2\sqrt{s}],
\qquad
E_R=\frac{h^2}{8ma^2}.
\label{eq:hoppingEstimate}
\end{equation}
For NaCs, taking \(m\simeq156.9\,{\rm amu}\) and \(a=532\,{\rm nm}\) gives \(E_R/h\simeq1.12\,{\rm kHz}\).
The full molecule-frame dipole moment is \(d=4.6\,{\rm D}\)~\cite{StevensonPRL2023,DengNatCommun2025}, but the coefficient of a dressed non-axisymmetric harmonic should be viewed as an effective dipole scale \(d_{\rm eff}\), set by the microwave dressing and generally much smaller than \(d\).
For an optical-lattice spacing \(a\), the corresponding nearest-neighbor dipolar scale is
\begin{equation}
\frac{V}{h}
\simeq
1.00\,{\rm kHz}
\left(\frac{d_{\rm eff}}{1\,{\rm D}}\right)^2
\left(\frac{532\,{\rm nm}}{a}\right)^3 .
\label{eq:Vestimate}
\end{equation}
Table~\ref{tab:feasibility} shows representative numbers.
Effective anisotropic moments of order \(0.18\)--\(0.30\,{\rm D}\), far below the full NaCs molecular-frame dipole moment, already place \(V/t\) in the range where the stripe lobe and its first-order closure occur in Fig.~\ref{fig:phase}.
The estimates are intended only as a feasibility scale; a quantitative proposal would require computing \(U_{\rm eff}\) for a chosen microwave dressing protocol and evaluating Eq.~\eqref{eq:wannierV} with the actual Wannier functions.

\begin{table}[tbp]
\caption{
\textbf{Representative NaCs lattice scales.}
Estimates use \(a=532\,{\rm nm}\), \(m=156.9\,{\rm amu}\), and the conversion chain in Eqs.~\eqref{eq:ad2Nature}, \eqref{eq:D2ad2}, \eqref{eq:latticeScaleConversion}, \eqref{eq:hoppingEstimate}, and \eqref{eq:Vestimate}.
The reference \(a_{d2}^{\rm ref}=1.3\times10^4a_0\) is obtained from the NaCs droplet dressing parameters of Ref.~\cite{ZhangNature2026}, for which \(a_{d2}=(5.31\times10^4a_0)\sin(2\xi)\).
Here \(\eta=a_{d2}/a_{d2}^{\rm ref}\), while \(d_{\rm eff}\) denotes the equivalent effective coefficient of the engineered anisotropic harmonic, not the full molecule-frame dipole moment.
The tabulated \(V/h\) values use the point-particle deep-lattice projection; harmonic Gaussian-Wannier smearing is estimated to renormalize the offsite far-field matrix elements at the \(\sim10\%\) level for these depths.
}
\label{tab:feasibility}
\begin{ruledtabular}
\begin{tabular}{cccccc}
\(s\) & \(\eta\) & \(d_{\rm eff}\) (D) & \(t/h\) (Hz) & \(V/h\) (Hz) & \(V/t\) \\
\hline
8 & 0.18 & 0.30 & 42.1 & 92 & 2.18 \\
10 & 0.12 & 0.25 & 25.5 & 61 & 2.40 \\
12 & 0.064 & 0.18 & 16.0 & 33 & 2.04 \\
12 & 0.078 & 0.20 & 16.0 & 40 & 2.48 \\
15 & 0.064 & 0.18 & 8.4 & 33 & 3.90 \\
\end{tabular}
\end{ruledtabular}
\end{table}

These representative points correspond to tunneling frequencies \(t/h\simeq8\)--\(42\,{\rm Hz}\) and interaction frequencies \(V/h\simeq33\)--\(92\,{\rm Hz}\).
Equivalently, \(h\times1\,{\rm Hz}=48\,{\rm pK}\), so the interaction scale is roughly \(1.6\)--\(4.4\,{\rm nK}\).
The values should be read as an order-of-magnitude reachability estimate rather than as an optimized parameter set; increasing the lattice spacing, choosing a different depth, or changing the dressing strength moves the system through the same \(V/t\) window.

\subsection{Experimental signatures}

These results suggest a concrete protocol for extending continuum microwave-shielded molecular gases into lattice quantum simulation.
One would first prepare a shielded molecular gas, load it into a two-dimensional square lattice, choose a microwave dressing configuration that produces a nonzero \(D_2\) harmonic, and use the lattice depth to tune \(t\) while the dressing fixes the overall scale and angular form of the long-range interaction.
Changing the total molecule number, or reading different radii of a weakly trapped cloud within a local-density approximation, then scans the filling.
A boxlike potential would be preferable for precision measurements of the first-order boundaries, because it reduces trap averaging near the stripe onset.
Operationally, the sequence is: tune into the target \(V/t\) range, image density profiles over repeated shots, fold the Fourier peaks into reciprocal-lattice families, and then inspect shot-resolved or spatial-bin-resolved distributions near the lobe closure.

The primary density signature is not a single preselected Bragg point, but a peak that remains in the axial \((q,0)\) family while its position changes with filling.
With site-resolved or sufficiently high-resolution in-situ images, this can be measured by Fourier transforming each density profile and then averaging the squared amplitude,
\begin{equation}
S_{\rm exp}(\mathbf q)
=
\frac{1}{N_s^2}
\left\langle
\left|
\sum_j
\left[n_j^{(\ell)}-\bar n^{(\ell)}\right]
e^{i\mathbf q\cdot\mathbf r_j}
\right|^2
\right\rangle_{\ell},
\label{eq:Sexp}
\end{equation}
where \(\ell\) labels experimental shots and \(\bar n^{(\ell)}\) removes the shot-dependent uniform background.
For nonzero reciprocal-lattice momenta in a uniform system this is equivalent to the site-normalized structure-factor convention used in Eq.~\eqref{eq:Sq}; in a trapped cloud it also helps suppress the slowly varying density envelope.
The resulting peaks can then be folded and summed over the same reciprocal-lattice families used in Fig.~\ref{fig:V3_family}.
This type of correlation analysis is already standard in dipolar Hubbard quantum simulators with magnetic atoms~\cite{SuNature2023}.
%For molecular experiments without single-site readout, the same ordering vector can be accessed through noise correlations after expansion or through weak Bragg probes of the density response.

The superfluid diagnostic requires a different measurement.
The QMC winding estimator is a thermodynamic stiffness and is not directly observed in an experiment.
Its closest experimental counterpart is phase coherence after release from the lattice, as seen in time-of-flight interference or related finite-range coherence probes~\cite{GreinerNature2002,SantraNatCommun2017,BigagliNature2024}.
More direct information on directional stiffness can be obtained from hydrodynamic-response measurements such as directional expansion or Bragg spectroscopy of sound modes along the two lattice axes~\cite{TaoPRL2023}.
The anisotropic-superfluid regime predicted here would appear as coherent matter-wave response with different effective stiffnesses along \(x\) and \(y\), before a static density peak becomes large.
The most convincing test would compare responses for perturbations along the two lattice axes while keeping the bare optical-lattice tunneling isotropic.

Near the stripe-lobe closure, the experimental message is especially practical.
An average over shots or over a trapped cloud can show simultaneous density modulation and residual coherence, which could be mistaken for a narrow supersolid window.
The hard-core simulations instead predict first-order switching.
Therefore the decisive measurement is shot-resolved or bin-resolved statistics: histograms of the family-resolved stripe structure factor, local density or particle number, and coherence contrast should be examined for broad multi-sector structure and sharpening near the boundary.
If controlled parameter sweeps are available, hysteresis between forward and reverse sweeps of the dressing strength or lattice depth, visible as different jump positions in \(S_x^\star\) and coherence contrast, would provide an additional first-order check.
A clean control is to rotate or reverse the anisotropic dressing component.
If the stripe direction follows this rotation while the optical lattice is unchanged, the ordering is tied to the engineered interaction geometry rather than to residual hopping anisotropy or imaging artifacts.
These proposed signatures connect naturally to recent progress on stable bosonic molecular gases, molecular Bose--Einstein condensation, microwave shielding, and dressed-state control in NaCs and related platforms~\cite{SchindewolfNature2022,LinPRX2023,StevensonPRL2023,BigagliNatPhys2023,BigagliNature2024,ZhangPRL2024Magic,StevensonPRL2024,KarmanPRXQ2025,WeiZhangPRXQ2025,ZhangNature2026}.

\section{Discussion and Conclusion}
\label{sec:discussion}
\label{sec:conclusion}

The results above show that the angular structure of the engineered, sign-changing interaction controls the phase diagram of hard-core bosons on a square lattice.
It drives the ground state from a compressible superfluid state at larger \(t/V\) to an axial stripe-solid state at smaller \(t/V\).
On the superfluid side, the same interaction geometry imprints directional stiffness anisotropy before static density order appears; inside the stripe solid, it leaves a filling-dependent stripe period within the axial \((q,0)\) family.
Near the stripe-lobe closure, averaged observables can simultaneously retain residual superfluid response and an enhanced stripe structure factor.
The measurement-resolved stripe structure-factor histograms show that this apparent coexistence reflects first-order switching between superfluid and stripe-solid states, rather than a distinct stable supersolid phase.
This makes the present model a clean lattice benchmark for programmable molecular interactions: it isolates the many-body consequences of a non-axisymmetric interaction geometry while deliberately omitting microscopic details of a specific dressing protocol.

In more conventional extended Bose--Hubbard models with predominantly repulsive or tilted dipolar interactions, the leading ordered states on square and triangular lattices are commonly checkerboard solids, supersolids, or stripes arising from a competition among density couplings at different distances and orientations~\cite{GoralPRL2002,MenottiPRL2007,CapogrossoPRL2010,PolletPRL2010,DanshitaPRL2009,TrefzgerJPB2011,BandyopadhyayPRA2019,ZhangNJP2015TiltedDipolar,AleksandrovaPRA2024,ZhangPRA2021TiltedDipolar,MaciaPRA2014,ZhangPRA2022Tilted3D}.
By contrast, the present interaction is sign changing already at the two-body level: it is repulsive along one lattice axis, attractive along the other, and nodal along the diagonals.
This angular structure directly lowers the interaction energy in the axial \((q,0)\) stripe family, rather than selecting order only through a balance among several repulsive density couplings.
It also separates interaction anisotropy from kinetic anisotropy.
The hopping is isotropic throughout this work, so both the larger-\(t/V\) anisotropic superfluid and the smaller-\(t/V\) stripe solid arise from the interaction kernel itself.

The same perspective distinguishes the present lattice problem from continuum tilted-dipole and molecular-gas settings, where roton softening, stripe supersolidity, droplets, vortex textures, self-bound gases, and high-superfluidity strongly correlated regimes are central themes~\cite{MaciaPRL2012,MaciaPRA2014,LuPRL2015,BombinPRL2017,NorciaNature2021,BlandPRL2022,RecatiNatRevPhys2023,CasottiNature2024,SchmidtPRR2022,AleksandrovaPRA2024,JinPRL2025,LangenPRL2025,CiardiPRL2025,WeiZhangPRXQ2025,ZhangNature2026}.
Those continuum studies establish microwave-shielded polar molecules as a strongly correlated quantum-matter platform.
Our work addresses the complementary lattice question: once the molecules are loaded into a square optical lattice, the interaction is projected onto discrete bonds and the many-body instability is organized by reciprocal-lattice families.
The key outcome is therefore not simply stripe order, but selection of an entire \((q,0)\) family whose preferred wave vector changes with filling.

The absence of a broad supersolid window in the present hard-core simulations has a simple physical interpretation.
A stripe supersolid would require static density modulation together with mobile particles or holes that retain phase coherence through the ordered background.
In the strict hard-core model, the stripe solid is stabilized by the attractive and repulsive directions of the non-axisymmetric long-range interaction.
Particle hopping competes with this density arrangement, and the absence of onsite number fluctuations strongly suppresses the winding response once stripe order is established.
The measurement-resolved structure-factor histograms near the lobe closure show that the apparent coexistence signal arises from first-order switching between superfluid and stripe-solid sectors, rather than from a distinct hard-core supersolid phase.
Further comparisons across system sizes can quantify this first-order behavior by showing how density, winding, and stripe-structure-factor jumps sharpen with \(L\).
Beyond the strict hard-core limit, a natural future direction is to retain a finite onsite repulsion or a microscopic short-range shielding core.
Such a soft-core extension would test whether additional onsite density fluctuations can create mobile defects inside the same interaction-selected stripe family, or whether the first-order closure found here remains robust.
Other natural extensions within the same hard-core setting are finite-temperature studies of stripe-family melting and microscopic matching to particular microwave-dressing schemes.
A microscopic matching program would start from a dressed two-body potential, integrate it over lattice Wannier functions to obtain \(U\), \(t\), and \(V_{ij}\), include the weak trap used in the experiment, and compare the predicted stripe-family peaks with measured density correlations~\cite{StevensonPRL2023,BigagliNatPhys2023,BigagliNature2024,ZhangPRL2024Magic,StevensonPRL2024,KarmanPRXQ2025,DengNatCommun2025,ZhangNature2026}.

In summary, path-integral quantum Monte Carlo simulations of this hard-core molecular-lattice benchmark show a progression from an anisotropic superfluid at larger \(t/V\) to an axial \((q,0)\) stripe-family solid at smaller \(t/V\).
The ordered state is best characterized not by one fixed ordering vector, but by an interaction-selected family whose internal wave vector reorganizes with filling; at the lobe closure, measurement-resolved stripe structure-factor histograms identify first-order sector switching rather than a stable supersolid window.

Overall, our results identify interaction geometry itself as an organizing principle for molecular lattice quantum matter.
They provide a transparent lattice benchmark for future studies of non-axisymmetric molecular interactions, and they suggest how engineered dipolar interactions may generate directional many-body order beyond the conventional axisymmetric dipolar paradigm.

\bibliography{molecule}

\end{document}